\documentclass[prl,balancelastpage,twocolumn,footinbib]{revtex4-1}
\usepackage{color,amsthm,amsmath,amsfonts,graphicx,bm,epsfig,float,siunitx,multirow,xr,enumerate,epstopdf,soul,ulem}
\usepackage[breaklinks=true,colorlinks=true,linkcolor=blue,urlcolor=blue,citecolor=blue]{hyperref}
\usepackage[T1]{fontenc}

\begin{document}

\title{$\mathcal{P} \mathcal{T}$-symmetric Feedback Induced Linewidth
Narrowing}
\author{Yuanjiang Tang$^{1}$, Chao Liang$^{1}$, Xin Wen$^{1}$, Weipeng Li$%
^{1}$, An-Ning Xu$^{1}$}
\author{Yong-Chun Liu$^{1,2}$}
\email{ycliu@tsinghua.edu.cn}
\affiliation{$^{1}$State Key Laboratory of Low-Dimensional Quantum Physics, Department of
Physics, Tsinghua University, Beijing 100084, China}
\affiliation{$^{2}$Frontier Science Center for Quantum Information, Beijing
100084, China}
\date{\today}

\begin{abstract}
Narrow linewidth is a
long-pursuing goal in precision measurement and sensing. We propose a
parity-time symmetric ($\mathcal{P}\mathcal{T}$-symmetric) feedback method to narrow
the linewidths of resonance systems. By using a quadrature
measurement-feedback loop, we transform a dissipative resonance
system into a $\mathcal{P}\mathcal{T}$-symmetric system. Unlike the
conventional $\mathcal{P}\mathcal{T}$-symmetric systems that typically
require two or more modes, here the $\mathcal{P}\mathcal{T}$-symmetric
feedback system contains only a single resonance mode, which greatly extends the scope
of applications. The method enables remarkable linewidth narrowing and
enhancement of measurement sensitivity. We illustrate the concept in a
thermal ensemble of atoms, achieving a $48$-fold narrowing of the magnetic
resonance linewidth. By applying the method in magnetometry, we realize a $22$%
-times improvement of the measurement sensitivity. This work opens the
avenue for studying non-Hermitian physics and high-precision measurements in
resonance systems with feedback.
\end{abstract}

\maketitle

Linewidth is one of the key factors that determine the performance of resonance systems, such as atoms, optical cavities and mechanical resonators. Especially, for precision
measurement and sensing, we always strive for a narrow linewidth to achieve
better measurement sensitivity. In various precision experiments such as
atomic magnetometry \cite{Kominis_subfemtotesla_2003,Budker_Optical_2007},
atomic gyroscopy \cite{Kornack_Nuclear_2005}, nuclear magnetic resonance
spectroscopy \cite{Suefke_Parahydrogen_2017}, exploration of dark matter
\cite{Jiang_Search_2021,Afach_Search_2021} and exotic forces \cite%
{Su_Search_2021a}, a very narrow linewidth enables one to detect extremely weak
signals. On the other hand, narrow linewidth also represents long coherence
time, which is beneficial for quantum storage and quantum information
processing. In order to reduce the linewidth, various methods have been
proposed, e.g., antirelaxation coating of vessel
walls \cite{Wu_Wall_2021,Balabas_Polarized_2010a,Zhao_Method_2008},
the spin-exchange-relaxation-free mechanism \cite%
{Happer_Spinexchange_1973,Allred_HighSensitivity_2002}, the nonlinear
magneto-optical rotation approach \cite{Budker_Resonant_2002a} and the coherent
population trapping scheme \cite{Scully_Highsensitivity_1992} in atomic systems.
However, these methods typically have specific and stringent requirements, e.g., highly
demanding fabrication, strict magnetic shielding or specific energy levels.

In recent years, parity-time ($\mathcal{P}\mathcal{T}$) symmetry has
attracted much interest, inspired by its unique property of exhibiting
real energy spectra with non-Hermitian Hamiltonians \cite%
{Bender_Real_1998,Feng_NonHermitian_2017a,El-Ganainy_NonHermitian_2018a,Ozdemir_Parity_2019a}%
. Much progress has been made in different systems, including optics
\cite{Makris_Beam_2008,Guo_Observation_2009,Ruter_Observation_2010a,Regensburger_Parity_2012b,Chang_Parity_2014,Peng_Parity_2014,Feng_Singlemode_2014},
atoms \cite{Hang_Symmetry_2013,Zhang_Observation_2016,Li_Observation_2019a,Ding_Experimental_2021a}, electronics \cite%
{Schindler_Experimental_2011a,Sun_Experimental_2014,Yang_Observation_2022},
nitrogen-vacancy centers \cite{Wu_Observation_2019}, optomechanics \cite%
{Jing_mathcalPT_2014,Lu_mathcalP_2015,Xu_Mechanical_2015,Schonleber_Optomechanical_2016,Zhang_phonon_2018},
acoustics \cite{Zhu_Symmetric_2014,Fleury_invisible_2015}, and microwave
systems \cite{Bittner_Symmetry_2012}, with potential in sensing \cite{Hodaei_Enhanced_2017a,Chen_Exceptional_2017,Lai_Observation_2019a,Xiao_Enhanced_2019,Kononchuk_Exceptionalpointbased_2022} . The study of $\mathcal{P}\mathcal{T}$
symmetry provides a powerful tool of engineering gain and loss and
controlling the system linewidth. However, previous realizations of $%
\mathcal{P}\mathcal{T}$ symmetry required two or more modes, which is not
applicable in a large variety of systems that contain only a single
resonance mode.

In this work we realize a novel type of $\mathcal{P}\mathcal{T}$-symmetric system by using a single resonance mode, which leads to
efficient and tunable linewidth narrowing. First, the effective gain is
realized by introducing feedback. Feedback is a fundamental component of modern
control theory \cite{Bechhoefer_Feedback_2005}, and has been widely employed
in quantum systems \cite{Wiseman_Quantum_1994} with applications in laser
cooling
\cite{Hopkins_Feedback_2003,Kleckner_Subkelvin_2006,Hamerly_Advantages_2012,Wilson_Measurementbased_2015,Rossi_Enhancing_2017,Guo_Feedback_2019,Tebbenjohanns_Cold_2019,Sommer_Partial_2019,vanderLaan_SubKelvin_2021,Whittle_Approaching_2021,Schmid_Coherent_2022,Koch_Feedback_2010,Behbood_Feedback_2013,Hush_Controlling_2013,Ivanov_Braggreflectionbased_2014,Schemmer_Monte_2017}%
, phase transition \cite%
{Kopylov_Timedelayed_2015,Hurst_Measurementinduced_2019,Hurst_Feedback_2020,Ivanov_FeedbackInduced_2020,Buonaiuto_Dynamical_2021,Munoz-Arias_Simulation_2020}%
, quantum state preparation \cite%
{Geremia_Deterministic_2006,Yanagisawa_Quantum_2006,Negretti_Quantum_2007,Sayrin_Realtime_2011,Zhou_Field_2012,Gajdacz_Preparation_2016,Lammers_Opensystem_2016,Wu_Cooling_2022}%
, quantum dynamical control\cite%
{Morrow_Feedback_2002,Steck_Quantum_2004,Vijay_Stabilizing_2012,Carmele_Single_2013,Vanderbruggen_Feedback_2013,Kohler_CavityAssisted_2017,Kroeger_Continuous_2020,Young_Feedbackstabilized_2021}%
, nonlinearity generation \cite%
{Jiang_Floquet_2021,Lloyd_Quantum_2000a,Liu_Feedbackinduced_2013,Munoz-Arias_Simulating_2020,Cuairan_Precision_2022}%
, and quantum amplification \cite{Yamamoto_Quantum_2016,Shimazu_Quantum_2021} as
well as generation of squeezing \cite%
{Thomsen_Spin_2002,Inoue_Unconditional_2013,Wade_Squeezing_2015,Cox_Deterministic_2016}
and entanglement \cite%
{Wang_Dynamical_2005,Carvalho_Controlling_2008,Riste_Deterministic_2013,Sudhir_Appearance_2017a}%
. Here we make use of positive feedback to transform loss into gain.
Second, we design a measurement-feedback loop in which
the two quadrature components are controlled separately,
which breaks the symmetry between the quadratures, so that the two
quadratures of a single resonance mode behave like two different modes with gain and loss, and
the parity operation represents the interchange between the two quadratures
of the same mode. Therefore, unlike the conventional $\mathcal{P}%
\mathcal{T}$-symmetric systems that typically contain two or more modes,
here the $\mathcal{P}\mathcal{T}$-symmetric feedback system requires only a
single resonance mode,
which greatly extends the application of $\mathcal{P}\mathcal{T}$ symmetry in a diversity of single resonance systems. Such
a method shows great advantage
in linewidth narrowing and sensitivity enhancement. Using an
experimental setup with a thermal atomic ensemble, we observe a $48$-fold
narrowing of the magnetic resonance linewidth, and achieve a $22$-times
improvement of the measurement sensitivity for magnetometry.

\begin{figure*}[tb]
\centering
\includegraphics[width=0.95\linewidth]{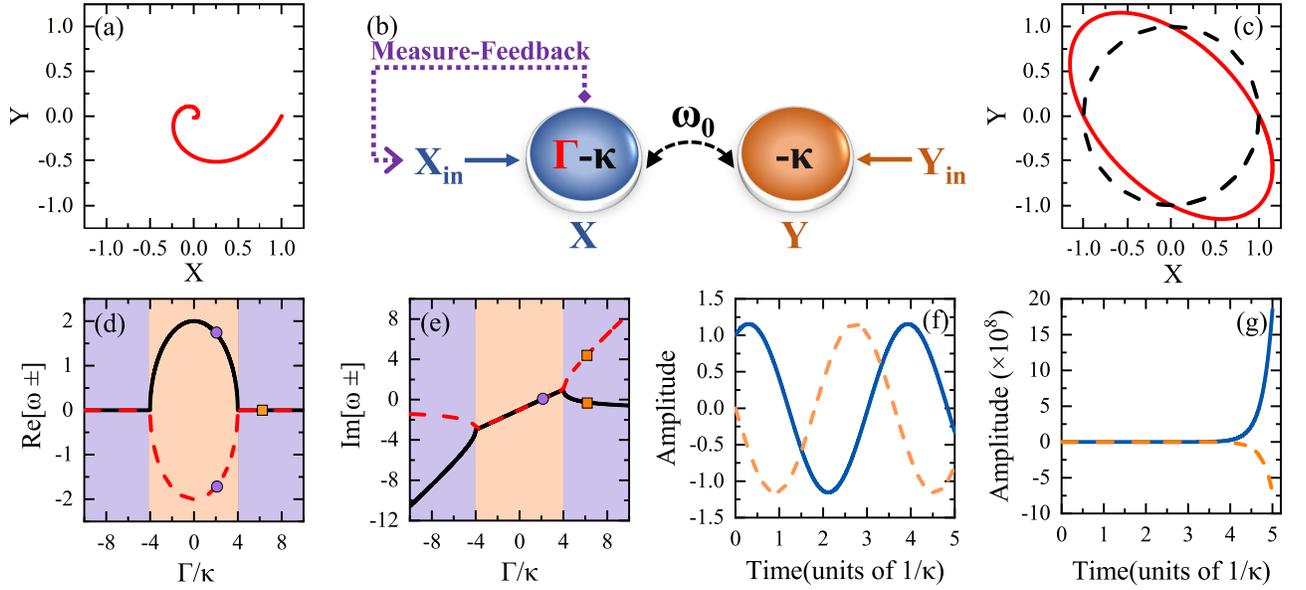}
\caption{(a) Typical trajectory of a dissipative resonance system in the
phase space. (b) Schematic diagram of the $\mathcal{P}\mathcal{T}$-symmetric
feedback system. (c) Phase-space trajectory of the $\mathcal{P}\mathcal{T}$%
-symmetric feedback system (red solid curve) and Hermitian resonance system
(black dashed curve). (d),(e) Real and imaginary parts of the eigenvalues
for $\Gamma/\protect\kappa$ varied from -10 to 10 and $\protect\omega_{0}/%
\protect\kappa=2$. The purple dots represent the case of %
$\Gamma=2 \kappa$, and the orange squares represent the case of $\Gamma=6 \kappa $, which are the eigenvalues of Figs.1(f) and (g), respectively. (f),(g) Typical time evolution of quadratures $X$ (blue solid curve) and $Y$
(orange dashed curve) in $\mathcal{P}\mathcal{T}$-symmetric phase ($%
\left\vert \Gamma \right\vert <2\protect\omega _{0}$) and symmetry-broken
phase ($\left\vert \Gamma \right\vert >2\protect\omega _{0}$). }
\label{fig1}
\end{figure*}

We consider a generic dissipative resonance system, which can be described
by the Hamiltonian $H=\omega _{0}a^{\dag }a$ and the corresponding quantum
Langevin equation $\dot{a}=(-i\omega _{0}-\kappa )a-\sqrt{2\kappa }a_{%
\mathrm{in}}$, where $a$ ($a^{\dag }$) is the annihilation (creation)
operator of the resonance mode, $\omega _{0}$ is the resonance frequency, $%
\kappa $ is the amplitude dissipation rate with the associate noise operator
being $a_{\mathrm{in}}$. The quadratures of the resonance mode are defined
as $X=(a+a^{\dag })/2$ and $Y=i(a^{\dag }-a)/2$, which correspond to the
bases in the phase space. $X_\mathrm{in}$ and $Y_\mathrm{in}$ correspond to the input quadratures. The equations of motion for the
quadratures are given by $\dot{X}=-\kappa X+\omega _{0}Y-\sqrt{2\kappa }X_{\mathrm{in}}$, $\dot{Y}=-\kappa
Y-\omega _{0}X-\sqrt{2\kappa }Y_{\mathrm{in}}.$ Because of the dissipation, the evolution trajectory in the
phase space is a spiral curve approaching the origin of coordinates, as
shown in Fig. \ref{fig1}(a). To construct a $\mathcal{P}\mathcal{T}$%
-symmetric Hamiltonian, as sketched in Fig. \ref{fig1}(b), we use a feedback
loop in which $X$ component is measured with the outcome feedback to input $X_{\mathrm{in}}$
component, i.e., $X_{\mathrm{in}}\rightarrow X_{\mathrm{in}}-\Gamma X/\sqrt{2\kappa}$, with $\Gamma $ being
the feedback parameter. Then the system dynamics is modified as
\begin{equation}
\left\{
\begin{array}{l}
\dot{X}=\left( \Gamma -\kappa \right) X+\omega _{0}Y-\sqrt{2\kappa }X_{\mathrm{in}}, \\
\dot{Y}=-\kappa Y-\omega _{0}X-\sqrt{2\kappa }Y_{\mathrm{in}}.%
\end{array}%
\right.
\end{equation}%
In the\ bases of quadratures with the vector $\Psi =\left( X\ Y\right) ^{T}$%
,\ the equations can be rewritten as $\dot{\Psi}=-iH_{\mathrm{eff}}\Psi $ with
\begin{equation}
H_{\mathrm{eff}}=\left(
\begin{array}{cc}
i\frac{\Gamma }{2} & i\omega _{0} \\
-i\omega _{0} & -i\frac{\Gamma }{2}%
\end{array}%
\right) +i\left( \frac{\Gamma }{2}-\kappa \right) \mathbf{I},  \label{eq3}
\end{equation}%
where $\mathbf{I}$ is the identity matrix. After dropping the identity matrix term that corresponds to a common gain or loss, the effective Hamiltonian is $%
\mathcal{P}\mathcal{T}$-symmetric. Here the parity operator $\mathcal{P}$ is
the Pauli operator $\sigma _{x}$ representing the interchange between the
two quadratures, and the time-reversal operator $\mathcal{T}$ denotes
complex conjugation operation. Therefore, the feedback transfers the
dissipative resonance system into a $\mathcal{P}\mathcal{T}$-symmetric
system, with equal gain and loss in two quadratures of the same resonance
mode. The key point is that the feedback breaks the symmetry between the
quadratures, and thus the two quadratures of a single resonance mode behave
like two different modes with gain and loss. As shown in Fig. \ref{fig1}(b), the effective
coupling strength between the quadratures is equal to the resonance
frequency $\omega _{0}$, as it is just the energy exchange frequency for
different components within the system.

For the Hermitian case without dissipation, the trajectory in the phase
space is a closed circle. However, in the $\mathcal{P}\mathcal{T}$-symmetric
case, the trajectory is squeezed into a oblique ellipse, as plotted in Fig. %
\ref{fig1}(c). In this case, the total effect of gain and loss for the $\pm
\pi /4$ quadratures $X_{\pm }=(X\pm Y)/\sqrt{2}$ are balanced, but the
couplings between $X_{+}$ and $X_{-}$ are rescaled as a
result of gain and loss, with $\dot{X}_{+}=-\left( \omega _{0}-\Gamma
/2\right) X_{-}$ and $\dot{X}_{-}=\left( \omega _{0}+\Gamma /2\right) X_{+}$%
. Therefore, the trajectory along $\pm \pi /4$ direction is squeezed
(stretched) by a factor of $1\pm \Gamma /(2\omega _{0})$, resulting in an
ellipse with an oblique angle of $\pi /4$.

The resultant $\mathcal{P}\mathcal{T}$-symmetric system possesses both $%
\mathcal{P}\mathcal{T}$-symmetric and $\mathcal{P}\mathcal{T}$%
-symmetry-broken phases, which can be tuned by the feedback parameter $%
\Gamma $. The eigenvalues of $H_{\mathrm{eff}}$ are ${\omega _{\pm }}%
=i\left( {\Gamma /2-\kappa }\right) \pm \sqrt{\omega _{0}^{2}-{\Gamma ^{2}/4}%
}$, whose real (imaginary) parts represent the resonance frequency
(dissipation rate) of the eigenmodes. In Fig. \ref{fig1}(d) and (e) we plot
the real and imaginary parts of the eigenvalues as functions of feedback
parameter $\Gamma $. When the feedback is weak ($\left\vert \Gamma
\right\vert <2\omega _{0}$), the real parts of the eigenvalues are opposite
to each other while the imaginary parts are equal, and the system is in the $%
\mathcal{P}\mathcal{T}$-symmetric phase. In this case the time evolutions of
the quadratures $X$ and $Y$ are still trigonometric functions, but the phase
difference is no longer $\pi /2$ [Fig. \ref{fig1}(f)], which is a result of
the phase advance or lag induced by the effective gain or loss. If the feedback is
strong enough ($\left\vert \Gamma \right\vert >2\omega _{0}$), the
eigenvalues are purely imaginary, indicating that the eigenmodes are no longer
harmonic modes. Then the amplitude of the quadratures increases exponentially
with time, as shown in Fig. \ref{fig1}(g).

\begin{figure}[tb]
\centering
\includegraphics[width=\linewidth]{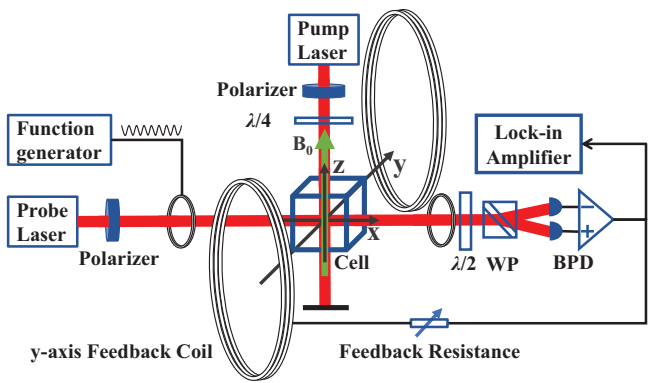}
\caption{Experimental setup with a thermal atomic ensemble. A $1$ cm$^{3}$
cube glass cell is rich in $^{133}$Cs, filled with $600$ Torr $^4{\rm{He}}$ and $20$ Torr ${{\rm{N}}_2}$ and
heated to $100$ $^{\circ }$C. The linearly polarized probe laser power is $%
25 $ $\protect\mu $W and frequency is $40$ GHz below the $F=4$ to $F^{\prime
}=2 $ transition of the $^{133}$Cs $D2$ line. The circularly polarized pump
laser power is $1$ mW and frequency is the $F=4$ to $F^{\prime }=3$
transition of the $^{133}$Cs $D1$ line. $B_0$, $z$-axis static magnetic field (green arrow), $\protect\lambda /4$, quarter
wave plate; $\protect\lambda /2$, half wavep late; WP, Wollaston prism; BPD,
balanced photodetector. The output signal of the BPD is applied to a loop
consisting of a feedback resistor and a $y$-axis feedback coil to achieve
magnetic field feedback.
The function generator is used to generate the weak driving magnetic field $B_x$.}
\label{fig2}
\end{figure}

We demonstrate the $\mathcal{P}\mathcal{T}$-symmetric feedback mechanism in
a thermal atomic ensemble, which is a typical example of magnetic resonance
system. The experimental setup is sketched in Fig. \ref{fig2}. A ensemble of
thermal cesium atoms is filled in a vapor cell, and the atoms can be
described by a collective spin with spin polarization $\mathbf{P}%
=(P_{x},P_{y},P_{z})$, where $P_{\mu =x,y,z}$ is the spin polarization
component along $\mu $ axis. A beam of circularly polarized laser
propagating along the $-z$ direction optically pumps the atomic ensemble to
polarize the collective spin. A static magnetic field of $B_{0}=2.2\ \mathrm{%
\mu T}$ is applied along $z$ axis, then the collective spin undergoes a
Larmor precession around $z$ axis, with Larmor frequency being $\omega
_{0}=\gamma B_{0}$, where $\gamma $ is the gyromagnetic ratio of the atom.
Thus, the transverse components $P_{x,y}$ oscillates
in the $xy$ plane, constituting a harmonic oscillator.
We measure the spin polarization component $P_{x}$ using a probe laser via
polarization homodyne detection, and the output signal
is then fed into a loop that includes a feedback
resistor and the $y$-axis feedback coil, which generates the feedback
magnetic field $B_{y}=-\Gamma _{\mathrm{FB}}P_{x}/\gamma $ with $\Gamma _{%
\mathrm{FB}}$ being the feedback factor. No noise processing of the signal is necessary because the signal-to-noise ratio in our experiments is large enough. In this case the feedback magnetic
field $B_{y}$ carries the information of the spin polarization component $%
P_{x}$,
which will result in $\mathcal{P}\mathcal{T}$-symmetric
feedback.

\begin{figure}[tb]
\centering
\includegraphics[width=\linewidth]{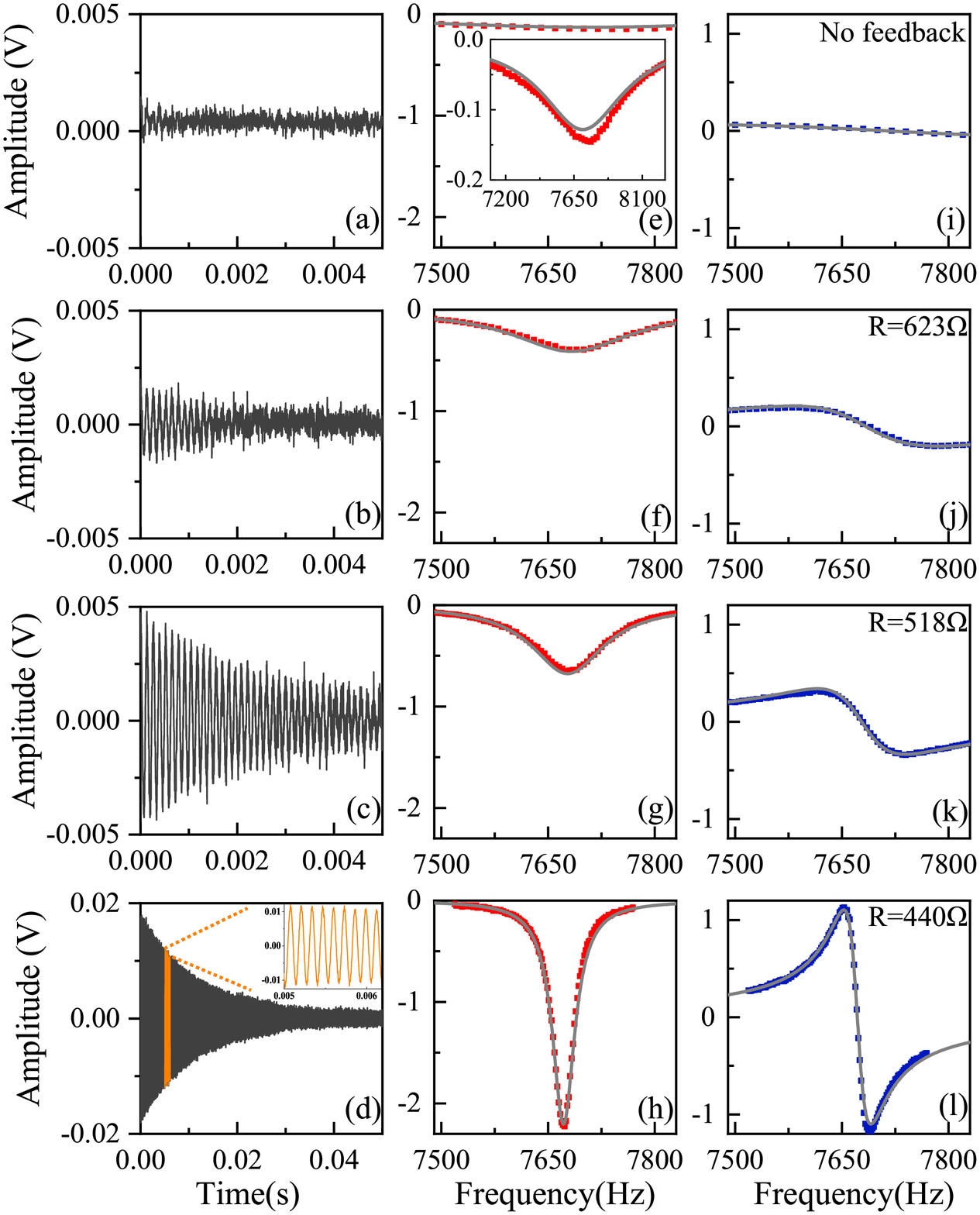}
\caption{From (a) to (e) are the output signals of the BPD after turning off
the driving magnetic field $B_{x}$. The inset in the (d) is a
zoomed-in view of the orange area. From (e) to (h) and (i) to (l) are the out
of phase and in phase output signals of the lock-in amplifier, respectively. The inset in the (e) shows a larger frequency range.
The scatters are experimental results and the gray solid curves are
theoretical results. From top to bottom, the feedback resistance decreases
with the values labeled in the figure.}
\label{fig3}
\end{figure}

When the feedback magnetic field $B_{y}$ is small compared with the static
magnetic field $B_{0}$, its effect on $P_{z}$ can be ignored, and $P_{z}$
remains equilibrium polarization $P_{0}$. Starting from the Bloch equations,
we obtain the simplified equations containing only two orthogonal components $%
(P_{x},P_{y})$ as
\begin{equation}
\left\{
\begin{array}{l}
\dot{P}_{x}=\left( \Gamma _{\mathrm{FB}}P_{0}-\frac{1}{T_{2}}\right)
P_{x}+\omega _{0}P_{y}, \\
\dot{P}_{y}=-\omega _{0}P_{x}-\frac{1}{T_{2}}P_{y},%
\end{array}%
\right.  \label{eq5}
\end{equation}%
where $T_{2}$ is the transverse relaxation time.
The dynamics can be effectively described by
\begin{equation}
H_{\mathrm{eff}}=\left(
\begin{array}{cc}
i\frac{P_{0}}{2}\Gamma _{\mathrm{FB}} & i\omega _{0} \\
-i\omega _{0} & -i\frac{P_{0}}{2}\Gamma _{\mathrm{FB}}%
\end{array}%
\right) +i\left( \frac{P_{0}}{2}\Gamma _{\mathrm{FB}}-\frac{1}{T_{2}}\right)
\mathbf{I},  \label{eq6}
\end{equation}%
which is equivalent to Eq. (\ref{eq3})
with $\kappa =1/T_{2}$ and $\Gamma =\Gamma _{\mathrm{FB}}P_{0}$. Therefore, the collective spin oscillator constitutes a $\mathcal{P}\mathcal{T}$-symmetric system.

Next we focus on the $\mathcal{P}\mathcal{T}$-symmetric phase ($\Gamma _{\mathrm{%
FB}}P_{0}<2\omega _{0}$) and show the ability of linewidth narrowing. The
imaginary part of the eigenvalues is $\Gamma _\mathrm{FB}P_{0}/2-1/T_{2}$, thus the linewidth $\Delta\omega _{\rm{FWHM}}$ is
\begin{equation}
\Delta\omega _{\rm{FWHM}}=\frac{2}{T_{2}}-\Gamma _{\mathrm{FB}}{P_{0}}.
\end{equation}
As the feedback factor $\Gamma _{\mathrm{FB}}$ increases, the
system dissipation keeps reducing, and the resonance linewidth keeps
narrowing. In our experiment, the feedback factor is inversely proportional
to the feedback resistance $(\Gamma _{\mathrm{FB}}\propto 1/R)$. Therefore,
our scheme enables flexible adjusting of the linewidth by changing the
feedback factor $\Gamma _{\mathrm{FB}}$ through the resistance $R$.

\begin{figure}[tb]
\centering
\includegraphics[width=0.95\linewidth]{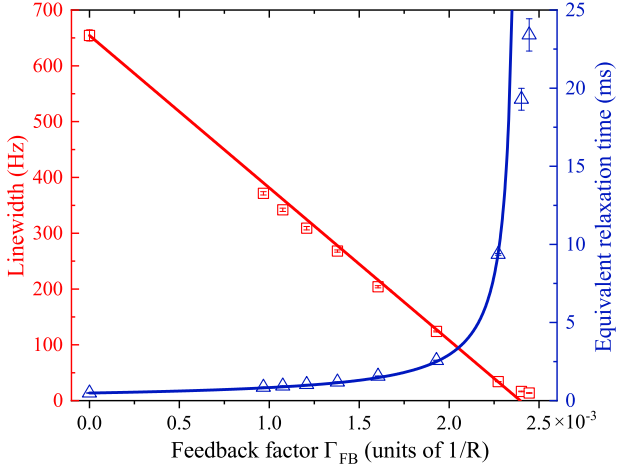}
\caption{Dependence of linewidth and equivalent relaxation time on the
feedback factor. The scatters are experimental results and the red and blue
solid line is the theoretical result.}
\label{fig4}
\end{figure}

In order to measure the resonance linewidth, we apply a weak
driving magnetic field $B_{x}=B_{1}\cos \left( \omega t\right) $ along $x$ axis, which corresponds to an additional term $\gamma B_{1}\cos \left(
\omega t\right) $ in the equation of $P_{y}$ in Eq. (\ref{eq5}). Then the
system undergoes forced oscillation, and we can use instantaneous drive with
sudden tuning off to obtain the system evolution dynamics in the time
domain, or use continuous drive with scanning frequency $\omega $ to obtain
the system response in the frequency domain. In Fig. \ref{fig3}, the time
domain signals (first column), frequency domain absorption signals (second
column) and dispersion signals (third column) are plotted. From top to
bottom, as the feedback resistance $R$ decreases (corresponding to the
increase of the feedback factor $\Gamma _{\mathrm{FB}}$), it shows clearly
that the oscillation lasts longer, and the absorption linewidth becomes
narrower, and the dispersion slope becomes sharper. The experimental results
are in good agreement with the theoretical predictions, where
the feedback delay has been taken into account (see Supplemental Material \cite{supp}).

The dependencies of linewidth and equivalent relaxation time on the feedback
factor are plotted in Fig. \ref{fig4}. In the experiment, we have observed
the reduction of the linewidth from $654$ to $13.6$ Hz, which is $48$ times narrower. The equivalent relaxation time increases from $0.486$ to
$23.4$ ms, which significantly extends the coherence time of the system.
Further improvement is limited by the stability of the present experimental
system, as it becomes more sensitive to the parameter variations
when the linewidth is very narrow. 

\begin{figure}[tb]
\centering
\includegraphics[width=\linewidth]{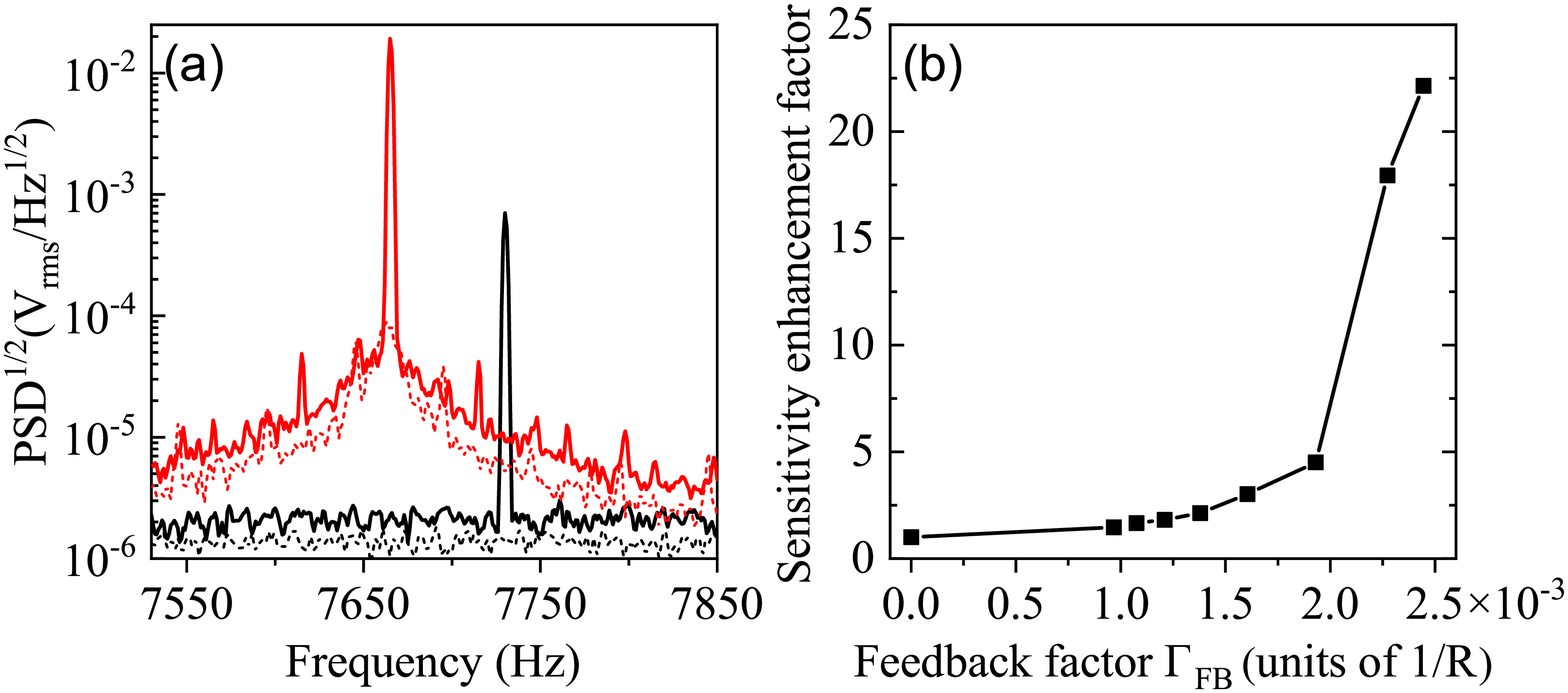}
\caption{(a) Square root of power spectral density (PSD$^{1/2}$) for
feedback resistance $R=409\ \Omega$ (red curves) and without feedback (black
curves). The solid (dashed) curves are the results in the presence (absence)
of driving magnetic field $B_{x}$. (b) Sensitivity enhancement factor of the
$M_{x}$ magnetometer as a function of the feedback factor.}
\label{fig5}
\end{figure}

The $\mathcal{P}\mathcal{T}$-symmetric feedback induced linewidth narrowing
method holds great potential for high-precision measurements. Our
experimental system can be directly used to improve the
measurement sensitivity of magnetic field, with the apparatus
similar to the $M_{x}$ magnetometer \cite%
{Groeger_highsensitivity_2006,Budker_Optical_2013}. When the driving magnetic field
is on resonance with the Larmor frequency
$\omega _{0}=\gamma B_{0}$, the spin polarization $%
P_{x}$ reaches its maximum value. Thus the magnitude of the magnetic field $%
B_{0}$ can be obtained by scanning the frequency of the driving magnetic
field. The measurement sensitivity of this $M_{x}$ magnetometer is
\cite{Groeger_highsensitivity_2006,Budker_Optical_2013,supp}
\begin{equation}
\delta B=k_F\frac{\Delta\omega _{\rm{FWHM}}}{\gamma }\frac{1}{S/N},  \label{eq9}
\end{equation}%
where $k_F=\frac{1}{2\sqrt2}$. To obtain the signal-to-noise ratio $S/N$, we measure the square root of the
power spectral density (PSD$^{1/2}$) by feeding the output of time domain signals into
the fast fourier transformation (FFT) spectrum analyzer (SR760).
As compared in Fig. \ref{fig5}(a), the signal with feedback is
significantly larger compared to that without feedback. As the background
noise also increases, the signal-to-noise ratio stays almost unchanged. The
shift of resonance frequency originates from the relaxation and the feedback
with delay (see Supplemental Material \cite{supp}), which does not affect the
measurement sensitivity for small changes of magnetic field. According to
Eq. (\ref{eq9}), we can obtain the dependence of sensitivity on the feedback
factor, as plotted in Fig. \ref{fig5}(b). The sensitivity of the $M_{x}$
magnetometer is enhanced up to $22$ times, with the linewidth narrowing
playing a significant role in this enhancement. Compared with the $48$-fold
reduction of the linewidth, this $22$-times enhancement of the measurement
sensitivity indicates some additional noise in the feedback process, which
may be overcome by further stabilizing the feedback loop.

In summary, we propose a $\mathcal{P}\mathcal{T}$-symmetric feedback method
in a general dissipative resonance system. By constructing a quadrature
measurement-feedback loop in which one quadrature component is measured with feedback,
a purely dissipative
resonance system can be transformed into a $\mathcal{P}\mathcal{T}$%
-symmetric system, with tunable $\mathcal{P}\mathcal{T}$-symmetric phase and
$\mathcal{P}\mathcal{T}$-symmetry-broken phase. Such a $\mathcal{P}\mathcal{T%
}$-symmetric system contains only a single resonance mode, without the
requirement of two or more modes, as the feedback breaks the symmetry
between the quadratures, and thus the two quadratures of a single resonance
mode behave like two different modes. The method finds important
applications in linewidth narrowing and enhancement of measurement
sensitivity. We demonstrate the proposal in a thermal atomic ensemble and
observe a $48$-fold narrowing of the magnetic resonance linewidth. By
applying the method in the $M_{x}$ magnetometer, we realize a $22$-times
enhancement of the magnetic field measurement sensitivity. It can also be
directly applied to other precision measurement experiments limited by
linewidth such as atomic gyroscopy. Our study provides a new perspective on
using feedback to construct $\mathcal{P}\mathcal{T}$-symmetric systems,
which form an excellent platform for studying non-Hermitian physics, with
broad applications in high-precision measurement and sensing.

\begin{acknowledgments}
This work is supported by the Key-Area Research and Development Program of
Guangdong Province (Grants No. 2019B030330001), the National Natural Science
Foundation of China (NSFC) (Grants No. 12275145, No. 92050110, No. 91736106, No. 11674390, and No. 91836302), and the National Key R\&D Program of China (Grants No. 2018YFA0306504).
\end{acknowledgments}

\end{document}


\title{Supplemental Material for ``$\mathcal{P} \mathcal{T}$-symmetric
feedback induced linewidth narrowing''}
\author{Yuanjiang Tang$^{1}$, Chao Liang$^{1}$, Xin Wen$^{1}$, Weipeng Li$%
^{1}$, An-Ning Xu$^{1}$}
\author{Yong-Chun Liu$^{1,2}$}
\email{ycliu@tsinghua.edu.cn}
\affiliation{$^{1}$State Key Laboratory of Low-Dimensional Quantum Physics, Department of
Physics, Tsinghua University, Beijing 100084, China}
\affiliation{$^{2}$Frontier Science Center for Quantum Information, Beijing
100084, China}
\date{\today }
\maketitle
\tableofcontents






\section{Theoretical model of feedback with delay}

\label{sec1}

In the $\mathcal{P}\mathcal{T}$-symmetric feedback method as described in
the main text, the $x$-component of the spin polarization $P_{x}$ is
measured, and then fed back to the $y$-axis magnetic field coil. In general,
there is typically a delay time $\tau $ between the feedback and the
measurement processes. Because of the delay, the magnetic field generated by
the feedback at time $t$ is proportional to the measured spin polarization
at $t-\tau $, which can be written as
\begin{equation}
B_{y}(t)=-\frac{\Gamma _{\mathrm{FB}}}{\gamma }P_{x}\left( t-\tau \right) ,
\end{equation}%
where $\Gamma _{\mathrm{FB}}$ is the feedback factor and $\gamma $ is the
gyromagnetic ratio of the atom.
When the delay time $\tau $ is much smaller than the characteristic time of the system mentioned below (i.e., the relaxation time $T_2$ and the oscillation period $2\pi/\omega_0$), we can use Taylor expansion and retain
the first-order term of $\tau $, yielding
\begin{equation}
P_{x}\left( t-\tau \right) =P_{x}(t)-\tau \dot{P}_{x}(t).
\end{equation}%
Now all the physical quantities are at time $t$, so the time variable $t$ is
omitted below.

The total magnetic field vector is
\begin{equation}
\mathrm{\mathbf{B}}=\left( B_{x},B_{y},B_{z}\right) =\left( B_{1}\cos
(\omega t),-\frac{\Gamma _{\mathrm{FB}}}{\gamma }\left( P_{x}-\tau \dot{P}%
_{x}\right) ,B_{0}\right) ,
\end{equation}%
where $B_{x}$ is the weak driving magnetic field applied to measure the
system properties, $B_{y}$ is the feedback magnetic field, and $B_{z}$ is
the static bias magnetic field. The dynamic of the spin polarization $%
\mathrm{\mathbf{P}}=(P_{x},P_{y},P_{z})$ can be described by the Bloch
equation
\begin{equation}
\left\{
\begin{array}{l}
\dot{P}_{x}=\gamma \left( P_{y}B_{z}-P_{z}B_{y}\right) -\frac{P_{x}}{T_{2}},
\\
\dot{P}_{y}=\gamma \left( -P_{x}B_{z}+P_{z}B_{x}\right) -\frac{P_{y}}{T_{2}},
\\
\dot{P}_{z}=\gamma \left( P_{x}B_{y}-P_{y}B_{x}\right) -\frac{P_{z}}{T_{1}}%
+R_{\mathrm{op}},%
\end{array}%
\right.   \label{eqS1}
\end{equation}%
where $T_{1}$ is the longitudinal relaxation time, $T_{2}$ is the transverse
relaxation time, and $R_{\mathrm{op}}$ is the rate of optical pumping.
Putting the expression of total magnetic field $\mathrm{\mathbf{B}}$ into
Eq.(\ref{eqS1}) gives
\begin{equation}
\dot{P}_{x}=\omega _{0}P_{y}+\Gamma _{\mathrm{FB}}P_{z}\left( P_{x}-\tau
\dot{P}_{x}\right) -\frac{P_{x}}{T_{2}},  \label{eqS2}
\end{equation}%
\begin{equation}
\dot{P}_{y}=-\omega _{0}P_{x}+\omega _{1}P_{z}\cos (\omega t)-\frac{P_{y}}{%
T_{2}},  \label{eqS3}
\end{equation}%
\begin{equation}
\dot{P}_{z}=-\Gamma _{\mathrm{FB}}P_{x}\left( P_{x}-\tau \dot{P}_{x}\right)
-P_{y}\omega _{1}\cos (\omega t)-\frac{P_{z}}{T_{1}}+R_{\mathrm{op}},
\label{eqS4}
\end{equation}%
where $\omega _{0}=\gamma B_{0}$, $\omega _{1}=\gamma B_{1}$.

In Eq.(\ref{eqS4}), when the feedback factor $\Gamma _{\mathrm{FB}}$ is
small compared with the optical pumping rate $R_{\mathrm{op}}$, the
nonlinear term $-\Gamma _{\mathrm{FB}}P_{x}\left( P_{x}-\tau \dot{P}%
_{x}\right) $ can be ignored. $B_{1}$ is also small as it is a weak magnetic
field applied to measure the system properties, so the driving term $%
-P_{y}\omega _{1}\cos (\omega t)$ can be neglected. Then $P_{z}$ reaches the
steady state, corresponding to the equilibrium polarization $P_{z}=P_{0}=R_{%
\mathrm{op}}T_{1}$, which keeps a constant in the experiment.

From Eq.(\ref{eqS2}) we obtain
\begin{equation}
\dot{P}_{x}=\frac{\Gamma _{\mathrm{FB}}P_{0}-\frac{1}{T_{2}}}{1+\Gamma _{%
\mathrm{FB}}P_{0}\tau }P_{x}+\frac{\omega _{0}}{1+\Gamma _{\mathrm{FB}%
}P_{0}\tau }P_{y}.  \label{eqS5}
\end{equation}%
From Eq.(\ref{eqS5}) we obtain the derivative as
\begin{equation}
\ddot{P}_{x}=\frac{\Gamma _{\mathrm{FB}}P_{0}-\frac{1}{T_{2}}}{1+\Gamma _{%
\mathrm{FB}}P_{0}\tau }\dot{P}_{x}+\frac{\omega _{0}}{1+\Gamma _{\mathrm{FB}%
}P_{0}\tau }\dot{P}_{y}.  \label{eqS6}
\end{equation}%
Inserting $\dot{P}_{y}$ given by Eq.(\ref{eqS3}) into Eq.(\ref{eqS6}), we
obtain
\begin{equation}
\ddot{P}_{x}=\frac{\Gamma _{\mathrm{FB}}P_{0}-\frac{1}{T_{2}}}{1+\Gamma _{%
\mathrm{FB}}P_{0}\tau }\dot{P}_{x}-\frac{\omega _{0}^{2}}{1+\Gamma _{\mathrm{%
FB}}P_{0}\tau }P_{x}-\frac{\omega _{0}}{\left( 1+\Gamma _{\mathrm{FB}%
}P_{0}\tau \right) T_{2}}P_{y}+\frac{\omega _{0}\omega _{1}P_{0}}{1+\Gamma _{%
\mathrm{FB}}P_{0}\tau }\cos \omega t.  \label{eqS7}
\end{equation}%
From Eq.(\ref{eqS5}), we obtain the expression of $P_{y}$ as
\begin{equation}
P_{y}=\frac{1}{\omega _{0}}\left[ \left( 1+\Gamma _{\mathrm{FB}}P_{0}\tau
\right) \dot{P}_{x}-\Gamma _{\mathrm{FB}}P_{x}P_{0}+\frac{P_{x}}{T_{2}}%
\right] .  \label{eqS8}
\end{equation}%
Putting it into Eq.(\ref{eqS7}), finally we obtain 
\begin{equation}
\ddot{P}_{x}+\left( \frac{1}{T_{2}}-\frac{\Gamma _{\mathrm{FB}}P_{0}-\frac{1%
}{T_{2}}}{1+\Gamma _{\mathrm{FB}}P_{0}\tau }\right) \dot{P}_{x}+\left[ \frac{%
\omega _{0}^{2}}{1+\Gamma _{\mathrm{FB}}P_{0}\tau }-\frac{\Gamma _{\mathrm{FB%
}}P_{0}-\frac{1}{T_{2}}}{\left( 1+\Gamma _{\mathrm{FB}}P_{0}\tau \right)
T_{2}}\right] P_{x}=\frac{\omega _{0}\omega _{1}P_{0}}{1+\Gamma _{\mathrm{FB}%
}P_{0}\tau }\cos \omega t.
\end{equation}%
%
This describes a typical forced harmonic oscillation with a driving force,
which can be abbreviated as
\begin{equation}
\ddot{P}_{x}+\Gamma _{\mathrm{eff}}\dot{P}_{x}+\omega _{\mathrm{eff}%
}^{2}P_{x}=\frac{\omega _{0}\omega _{1}P_{0}}{1+\Gamma _{\mathrm{FB}%
}P_{0}\tau }\cos \omega t,  \label{eqS10}
\end{equation}%
where
\begin{eqnarray}
\Gamma _{\mathrm{eff}} &=&\frac{1}{T_{2}}-\frac{\Gamma _{\mathrm{FB}}P_{0}-%
\frac{1}{T_{2}}}{1+\Gamma _{\mathrm{FB}}P_{0}\tau },  \label{Gamma} \\
\omega _{\mathrm{eff}} &=&\sqrt{\frac{\omega _{0}^{2}}{1+\Gamma _{\mathrm{FB}%
}P_{0}\tau }-\frac{\Gamma _{\mathrm{FB}}P_{0}-\frac{1}{T_{2}}}{\left(
1+\Gamma _{\mathrm{FB}}P_{0}\tau \right) T_{2}}}.  \label{omega}
\end{eqnarray}

In order to calculate $P_{x}$, we can set $P_{x}=\mathrm{Re}[\tilde{P}%
_{x}e^{i\omega t}]$ and put it into Eq.(\ref{eqS10}), then $\tilde{P}_{x}$
satisfies the equation
\begin{equation}
( -\omega ^{2}+i\Gamma _{\mathrm{eff}}\omega +\omega _{\mathrm{eff}}^{2})%
\tilde{P}_{x}e^{i\omega t}=\frac{\omega _{0}\omega _{1}P_{0}}{%
1+\Gamma _{\mathrm{FB}}P_{0}\tau }e^{i\omega t}  \label{eqS11}
\end{equation}%
The solution of Eq.(\ref{eqS11}) is
\begin{equation}
\tilde{P}_{x}=\frac{\frac{\omega _{0}\omega _{1}P_{0}}{1+\Gamma _{\mathrm{FB}%
}P_{0}\tau }}{\omega _{\mathrm{eff}}^{2}-\omega ^{2}+i\Gamma _{\mathrm{eff}%
}\omega }=\frac{\frac{\omega _{0}\omega _{1}P_{0}}{1+\Gamma _{\mathrm{FB}%
}P_{0}\tau }\left( \omega _{\mathrm{eff}}^{2}-\omega ^{2}\right) }{\left(
\omega _{\mathrm{eff}}^{2}-\omega ^{2}\right) ^{2}+\Gamma _{\mathrm{eff}%
}^{2}\omega ^{2}}-i\frac{\frac{\omega _{0}\omega _{1}P_{0}}{1+\Gamma _{%
\mathrm{FB}}P_{0}\tau }\Gamma _{\mathrm{eff}}\omega }{\left( \omega _{%
\mathrm{eff}}^{2}-\omega ^{2}\right) ^{2}+\Gamma _{\mathrm{eff}}^{2}\omega
^{2}}.  \label{eqS12}
\end{equation}%
So we obtain
\begin{equation}
P_{x}=\mathrm{Re}\left[ \tilde{P}_{x}e^{i\omega t}\right] =\frac{\frac{%
\omega _{0}\omega _{1}P_{0}}{1+\Gamma _{\mathrm{FB}}P_{0}\tau }\left( \omega
_{\mathrm{eff}}^{2}-\omega ^{2}\right) }{\left( \omega _{\mathrm{eff}%
}^{2}-\omega ^{2}\right) ^{2}+\Gamma _{\mathrm{eff}}^{2}\omega ^{2}}\cos
(\omega t)+\frac{\frac{\omega _{0}\omega _{1}P_{0}}{1+\Gamma _{\mathrm{FB}%
}P_{0}\tau }\Gamma _{\mathrm{eff}}\omega }{\left( \omega _{\mathrm{eff}%
}^{2}-\omega ^{2}\right) ^{2}+\Gamma _{\mathrm{eff}}^{2}\omega ^{2}}\sin
(\omega t).  \label{eqS13}
\end{equation}%
When near resonance ($\omega \approx \omega _{\mathrm{eff}}$), the above
equation can be simplified as
\begin{equation}
P_{x}\simeq \frac{\frac{\omega _{0}\omega _{1}P_{0}}{2\omega _{\mathrm{eff}%
}\left( 1+\Gamma _{\mathrm{FB}}P_{0}\tau \right) }\left( \omega _{\mathrm{eff%
}}-\omega \right) }{\left( \omega _{\mathrm{eff}}-\omega \right) ^{2}+\frac{%
\Gamma _{\mathrm{eff}}^{2}}{4}}\cos (\omega t)+\frac{\frac{\omega _{0}\omega
_{1}P_{0}}{4\omega _{\mathrm{eff}}\left( 1+\Gamma _{\mathrm{FB}}P_{0}\tau
\right) }\Gamma _{\mathrm{eff}}}{\left( \omega _{\mathrm{eff}}-\omega
\right) ^{2}+\frac{\Gamma _{\mathrm{eff}}^{2}}{4}}\sin (\omega t).
\label{Px}
\end{equation}
The spin polarization $P_{x}$ contains two orthogonal components, $\cos
\left( \omega t\right) $ term and $\sin \left( \omega t\right) $ term, which
correspond to the dispersion and absorption properties, respectively. In
Fig. 3 of the main text, the curves are fitted with the two components in
Eq.(\ref{Px}).

When the feedback delay is relatively small with $\Gamma _{\mathrm{FB}%
}P_{0}\tau \ll 1$, from Eq.(\ref{Gamma}) and Eq.(\ref{omega}), we can obtain
\begin{eqnarray}
\Gamma _{\mathrm{eff}} &\simeq &\frac{2}{T_{2}}-\Gamma _{\mathrm{FB}}P_{0},
\\
\omega _{\mathrm{eff}} &\simeq &\sqrt{\omega _{0}^{2}-\Gamma _{\mathrm{FB}%
}P_{0}\tau \omega _{0}^{2}-\frac{1}{T_{2}^{2}}}.  \label{omega2}
\end{eqnarray}
Therefore, the impact of delay time $\tau $ on the linewidth is negligible,
while its impact on the resonance frequency is relatively large. In Fig.
5(a) of the main text, the resonance shift is determined by Eq.(\ref{omega2}).

\subsection{Theoretical model of feedback with delay and noise}

\label{sec1A}

In the original feedback model, we assumed that the total noise during the measurement is $N$, so the measured spin polarization $P_x$ should be added with this noise, and then fed back to the $y$-axis magnetic field $B_y$, so
\begin{equation}
	B_{y}(t)=-\frac{\Gamma\ _{\mathrm{FB}}}{\gamma\ }\left(P_{x}\left(\ t-\tau\ \right)+N\right).
\end{equation}
Considering the noise, the total magnetic field feedback is
\begin{equation}
	\mathrm{\mathbf{B}}=\left( B_{x},B_{y},B_{z}\right) =\left( B_{1}\cos
	(\omega t),-\frac{\Gamma _{\mathrm{FB}}}{\gamma }\left( P_{x}-\tau \dot{P}%
	_{x}+N\right) ,B_{0}\right),
\end{equation}
Putting the expression of total magnetic field $\mathrm{\mathbf{B}}$ into
Eq.(\ref{eqS1}) gives
\begin{equation}
	\dot{P}_{x}=\omega _{0}P_{y}+\Gamma _{\mathrm{FB}}P_{z}\left( P_{x}-\tau
	\dot{P}_{x}+N\right) -\frac{P_{x}}{T_{2}},  \label{eqS24}
\end{equation}%
\begin{equation}
	\dot{P}_{y}=-\omega _{0}P_{x}+\omega _{1}P_{z}\cos (\omega t)-\frac{P_{y}}{%
		T_{2}},  \label{eqS25}
\end{equation}%
\begin{equation}
	\dot{P}_{z}=-\Gamma _{\mathrm{FB}}P_{x}\left( P_{x}-\tau \dot{P}_{x}+N\right)
	-P_{y}\omega _{1}\cos (\omega t)-\frac{P_{z}}{T_{1}}+R_{\mathrm{op}}.
	\label{eqS26}
\end{equation}%
In Eq.(\ref{eqS26}), $B_{1}$ is small as it is a weak magnetic
field applied to measure the system properties, so the transverse spin components produced by the driving field $B_1$ satisfy $P_x,P_y\ll1$. Thus, the driving term $%
-P_{y}\omega _{1}\cos (\omega t)$ can be neglected. When the feedback term $\Gamma _{\mathrm{FB}}P_x$ is much small than $P_{z}/T_{1}$ and $R_{\mathrm{op}}$, the nonlinear term with noise $-\Gamma _{\mathrm{FB}}P_{x}\left( P_{x}-\tau \dot{P}%
_{x}+N\right) $ can be ignored. Then $P_{z}$ reaches the
steady state, corresponding to the equilibrium polarization $P_{z}=P_{0}=R_{%
	\mathrm{op}}T_{1}$,
which keeps a constant in the experiment.

We deal with Eq.(\ref{eqS24}) and Eq.(\ref{eqS25}) in the same way as we do with the noiseless feedback equations Eq.(\ref{eqS2}) and Eq.(\ref{eqS3}),%
we can get
\begin{equation}
	\ddot{P}_{x}+\Gamma _{\mathrm{eff}}\dot{P}_{x}+\omega _{\mathrm{eff}%
	}^{2}P_{x}=\frac{\omega _{0}\omega _{1}P_{0}}{1+\Gamma _{\mathrm{FB}%
		}P_{0}\tau }\cos \omega t
	+\frac{\Gamma _{\mathrm{FB%
		}}P_{0}N}{\left( 1+\Gamma _{\mathrm{FB}}P_{0}\tau \right)
		T_{2}},  \label{eqS27}
\end{equation}%
where
\begin{eqnarray}
	\Gamma _{\mathrm{eff}} &=&\frac{1}{T_{2}}-\frac{\Gamma _{\mathrm{FB}}P_{0}-%
		\frac{1}{T_{2}}}{1+\Gamma _{\mathrm{FB}}P_{0}\tau },  \label{Gamma2} \\
	\omega _{\mathrm{eff}} &=&\sqrt{\frac{\omega _{0}^{2}}{1+\Gamma _{\mathrm{FB}%
			}P_{0}\tau }-\frac{\Gamma _{\mathrm{FB}}P_{0}-\frac{1}{T_{2}}}{\left(
			1+\Gamma _{\mathrm{FB}}P_{0}\tau \right) T_{2}}}.
\end{eqnarray}
Due to the consideration of noise, Eq.(\ref{eqS27}) has an additional noise term on the right side of the equation compared to Eq.(\ref{eqS10})
It can be found that the noise term only corrects the driving term of the damped harmonic oscillator equation and does not affect the dissipation coefficient. Therefore, the noise term of the feedback does not affect the linewidth.

Next, we analyze the correction amplitude $\varepsilon$ of the driver term generated by the noise term. Comparing the coefficients of the two terms, we get
\begin{equation}
	\varepsilon=\frac
	{\frac{\Gamma _{\mathrm{FB%
			}}P_{0}N}{\left( 1+\Gamma _{\mathrm{FB}}P_{0}\tau \right)
			T_{2}}}{\frac{\omega _{0}\omega _{1}P_{0}}{1+\Gamma _{\mathrm{FB}%
			}P_{0}\tau}}=\frac{\Gamma _{\mathrm{FB}}}{\omega_{0}}\times\frac{N}{\omega_{1}T_2}.
\label{eqS30}
\end{equation}
In our experiment, the minimum effective linewidth is $\Gamma_{\mathrm{eff}}=2\pi\times10$ Hz, the Larmor frequency $\omega_0=\gamma B_0=2\ \pi\times0.35\mathrm{\ Hz/nT}\times2200\mathrm{\ nT}=2\ \pi\times7.7\mathrm{\ kHz}$, and the relaxation time $T_2=0.486\mathrm{\ ms}$, so
\begin{equation}
	\frac{\Gamma _{\mathrm{FB}}}{\omega_{0}}\approx\frac{2/{T_2}}{\omega_{0}}\approx\frac{2/{(0.486\ \text{ms})}}{2\pi\times7.7\  \text{kHz}}\approx0.085.
\label{eqS31}
\end{equation}
And our signal-to-noise ratio ($S/N$) is about 500. From Eq.(\ref{Px}) we can get the magnitude of the signal in the near resonance condition $\omega=\omega_{\text{eff}}\approx\omega_{0}$ and the delay $\tau$ is small,
\begin{equation}
S = \left| {{P_x}} \right| \simeq \sqrt {{{\left( {\frac{{\frac{{{\omega _0}{\omega _1}{P_0}}}{{2{\omega _{{\rm{eff}}}}\left( {1 + {\Gamma _{{\rm{FB}}}}{P_0}\tau } \right)}}\left( {{\omega _{{\rm{eff}}}} - \omega } \right)}}{{{{\left( {{\omega _{{\rm{eff}}}} - \omega } \right)}^2} + \frac{{\Gamma _{{\rm{eff}}}^2}}{4}}}} \right)}^2} + {{\left( {\frac{{\frac{{{\omega _0}{\omega _1}{P_0}}}{{4{\omega _{{\rm{eff}}}}\left( {1 + {\Gamma _{{\rm{FB}}}}{P_0}\tau } \right)}}{\Gamma _{{\rm{eff}}}}}}{{{{\left( {{\omega _{{\rm{eff}}}} - \omega } \right)}^2} + \frac{{\Gamma _{{\rm{eff}}}^2}}{4}}}} \right)}^2}}  = \frac{{\frac{{{\omega _0}{\omega _1}{P_0}}}{{2{\omega _{{\rm{eff}}}}\left( {1 + {\Gamma _{{\rm{FB}}}}{P_0}\tau } \right)}}}}{{\sqrt {{{\left( {{\omega _{{\rm{eff}}}} - \omega } \right)}^2} + \frac{{\Gamma _{{\rm{eff}}}^2}}{4}} }} \simeq \frac{{{\omega _1}}}{{{\Gamma _{{\rm{eff}}}}}}.
\end{equation}
So the signal-to-noise ratio $S/N$ is
\begin{equation}
	\frac{S}{N}=\frac{\omega_{1}/\Gamma_{\mathrm{eff}}}{N}\approx500,
\end{equation}
and we can obtain that
\begin{equation}
	\frac{\omega_{1}}{N}\approx500\times(2\pi\times10)\ \text{Hz}\approx31\ \text{kHz}.
	\label{eqS34}
\end{equation}
Putting Eq.(\ref{eqS31}) and (\ref{eqS34}) to the expression of $\varepsilon$, Eq.(\ref{eqS30}),
\begin{equation}
	\varepsilon\approx0.085\times\frac{1}{31\ \text{kHz}\times0.486 \text{ms}}\approx0.56\%.%
\end{equation}
It can be found that under our experimental conditions, the correction amplitude brought by noise is much smaller than that of the driving term. To be precise, because the signal-to-noise ratio of our experiment is large enough, we can safely ignore the effect of noise.

\section{$\mathcal{P} \mathcal{T}$-symmetric feedback in the optical cavity}
Optical cavity is widely used in precision measurement experiments, and our $\mathcal{P} \mathcal{T}$-symmetric feedback scheme can also be applied to the system of optical cavity to realize the linewidth narrowing.
\begin{figure}[b]
	\centering
	\includegraphics[width=0.8\linewidth]{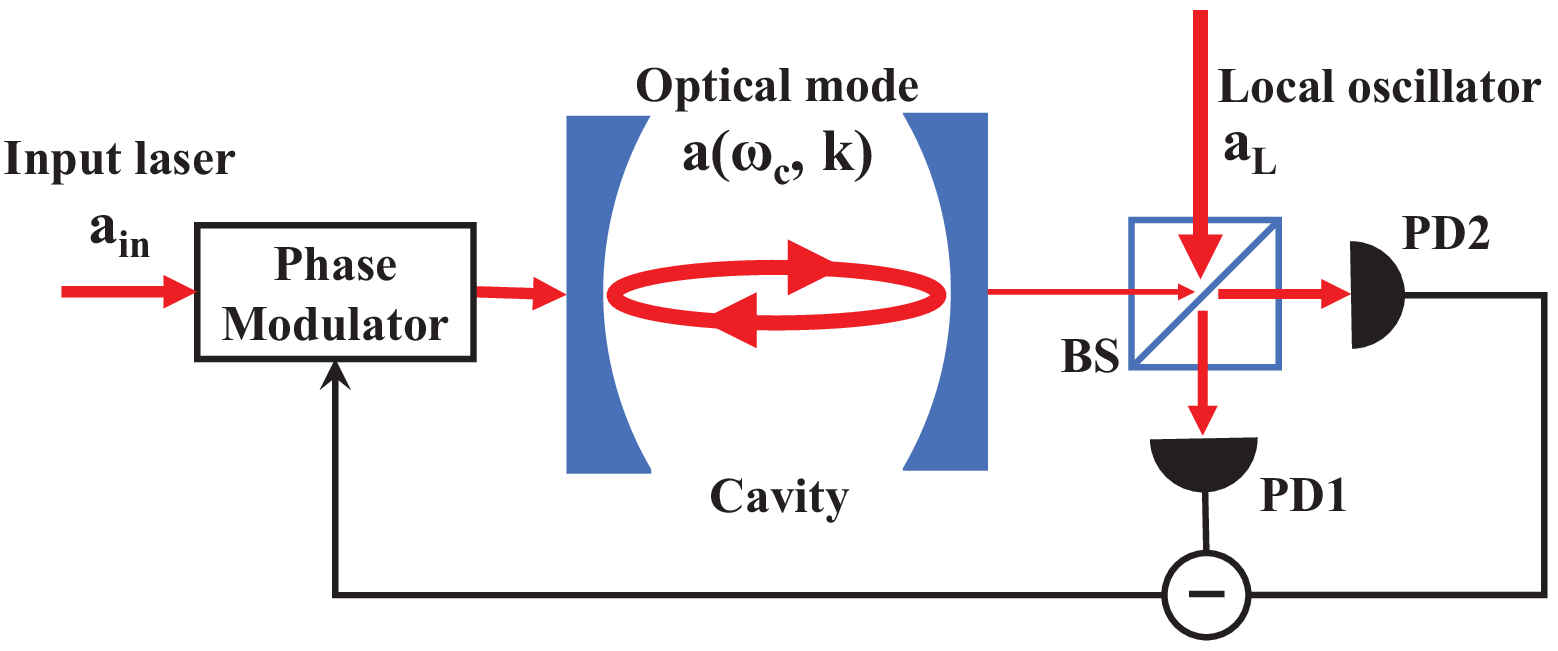}
	\caption{Schematic diagram of an optical cavity using PT feedback laser and pump laser used in the experiment. BS, beam splitter; PD, photodetector.}
	\label{figS1}
\end{figure}
As shown in the Fig.\ref{figS1}, the laser is incident into the optical cavity through the phase modulator. We detect the output laser by homodyne detection, and then the detection signal is fed back to the phase modulator to modulate the phase of the input optical field. $a\left(a^\dag\right)$ is the annihilation (creation) operator of the light field in the Fabry-Perot cavity, satisfying the Hamiltonian $H=(\omega_c-\omega_L)a^\dag a=-\mathrm{\Delta}a^\dag a$, where $\omega_c$ is the resonance frequency and $\omega_L$ is the frequency of the light field, $\mathrm{\Delta}$ is the detuning. So the Langevin equation for the light field in the cavity is $\dot{a}=\left(i\mathrm{\Delta}-\kappa\right)a-\sqrt{2\kappa_{\mathrm{in}}}a_{\mathrm{in}}$, where $\kappa_{\mathrm{in}}$ is the loss rate associated with the input coupling, $\kappa$ is the total loss rate. Since the annihilation operator $a$ can be written as two orthogonal components $a=X+iY$, the equation satisfied by the two orthogonal components X and Y is $\dot{X}+i\dot{Y}=\left(i\mathrm{\Delta}-\kappa\right)\left(X+iY\right)-\sqrt{\kappa_{\mathrm{in}}}\left(X_{\mathrm{in}}+iY_{\mathrm{in}}\right)$, which can be rewritten as
\begin{equation}
	\left\{\begin{array}{l}
		\dot{X}=-\kappa X-\Delta Y-\sqrt{2\kappa_{\mathrm{in}}} X_{\mathrm{in}} \\
		\dot{Y}=-\kappa Y+\Delta X-\sqrt{2\kappa_{\mathrm{in}}} Y_{\mathrm{in}}
	\end{array}\right..
\label{dotXdotY}
\end{equation}
Next, we introduce the measurement-feedback scheme. According to the input-output relation, the transmission light field satisfies $a_{\mathrm{out}}=\sqrt{\kappa_{\mathrm{in}}}a$. We use strong local oscillating light $a_L$ and transmission light interference for homodyne detection, and the light intensity of the two channels after splitting through the beam splitter satisfies
\begin{equation}
	\left\{\begin{array}{l}
		I_1 \propto\left(a_L^{\dagger}+a^{\dagger}\right)\left(a_L+a\right)=\left|a_L\right|^2+a^{\dagger} a+\left|a_L\right|\left(e^{-i \theta} a+e^{i \theta} a^{\dagger}\right) \\
		I_2 \propto\left(a_L^{\dagger}-a^{\dagger}\right)\left(a_L-a\right)=\left|a_L\right|^2+a^{\dagger} a-\left|a_L\right|\left(e^{-i \theta} a+e^{i \theta} a^{\dagger}\right)
	\end{array}\right.,
\end{equation}
where $\alpha_L=\left|\alpha_L\right|e^{i\theta}$, $\theta$ is the phase of the local light. The intensity difference between the two channels is
\begin{equation}
	I_1-I_2\propto2\left|a_L\right|\left(e^{-i\theta}a+e^{i\theta}a^\dag\right)=4\left|a_L\right|\left(X \cos{\theta}+Y \sin{\theta}\right).
\end{equation}
We can define
\begin{equation}
	Q_\theta=4\left|a_L\right|\left(X \cos{\theta}+Y \sin{\theta}\right),
\label{Qtheta}
\end{equation}
so we obtain $I_1-I_2\propto Q_\theta$.

The signal from the photodetector is fed back to a phase modulator (e.g., electro-optic modulator, EOM) to add an additional phase $\phi$ to the input laser $a_{\mathrm{in}}\rightarrow a_{\mathrm{in}}e^{i\phi}$, where $\phi\propto I_1-I_2$. Let $\eta$ be the total conversion efficiency, including the photodetector and phase modulator conversion efficiency, we can obtain $\phi=\eta Q_\theta$. When $\phi$ is small, we use the Taylor expansion
\begin{equation}
	a_{\mathrm{in}}e^{i\phi}\approx a_{\mathrm{in}}\left(1+i\phi\right)=\left(X_{\mathrm{in}}+iY_{\mathrm{in}}\right)\left(1+i\eta Q_\theta\right)=X_{\mathrm{in}}-\eta Y_{\mathrm{in}}Q_\theta+i\left(Y_{\mathrm{in}}+\eta X_{\mathrm{in}}Q_\theta\right)
\end{equation}
So the two orthogonal components of the input light are
\begin{equation}
\left\{ \begin{array}{l}
	X_{{\rm{in}}}^\prime  = {X_{{\rm{in}}}} - \eta {Y_{{\rm{in}}}}{Q_\theta }\\
	Y_{{\rm{in}}}^\prime  = {Y_{{\rm{in}}}} + \eta {X_{{\rm{in}}}}{Q_\theta }
\end{array} \right..
\label{xinyin}
\end{equation}
Putting Eq.(\ref{xinyin}) into Eq.(\ref{dotXdotY}), finally we obtain that the equation of the two orthogonal components of the optical field in the cavity is
\begin{equation}
	\left\{ \begin{array}{l}
		\dot X =  - \kappa X - \Delta Y - \sqrt {2{\kappa _{{\rm{in}}}}} \left( {{X_{{\rm{in}}}} - \eta {Y_{{\rm{in}}}}{Q_\theta }} \right)\\
		\dot Y =  - \kappa Y + \Delta X - \sqrt {2{\kappa _{{\rm{in}}}}} \left( {{Y_{{\rm{in}}}} + \eta {X_{{\rm{in}}}}{Q_\theta }} \right)
	\end{array} \right.
\label{dotXdotY2}
\end{equation}
It is worth noting that the feedback breaks the symmetry of the two orthogonal components.

When the adjustable phase of the local oscillation light satisfies $\theta=0$. From Eq.(\ref{Qtheta}), we know $Q_{\theta=0}=4\left|a_L\right|X$. And putting it into Eq.(\ref{dotXdotY2}), we obtain
\begin{equation}
	\left\{ \begin{array}{l}
		\dot X =  - \kappa X - \Delta Y - \sqrt {2{\kappa _{{\rm{in}}}}} \left( {{X_{{\rm{in}}}} - 4\left| {{a_L}} \right|\eta {Y_{{\rm{in}}}}X} \right) = \left( { - \kappa  + 4\sqrt {2{\kappa _{{\rm{in}}}}}\eta \left| {{a_L}} \right|{Y_{{\rm{in}}}}} \right)X - \Delta Y- \sqrt {2{\kappa _{{\rm{in}}}}} {{X_{{\rm{in}}}}}\\
		\dot Y =  - \kappa Y + \Delta X - \sqrt {2{\kappa _{{\rm{in}}}}} \left( {{Y_{{\rm{in}}}} + 4\left| {{a_L}} \right|\eta {X_{{\rm{in}}}}X} \right) =  - \kappa Y + \left( \Delta-4\sqrt {2{\kappa _{{\rm{in}}}}}\eta \left| {{a_L}} \right|{X_{{\rm{in}}}} \right) X - \sqrt {2{\kappa _{{\rm{in}}}}} {Y_{{\rm{in}}}}.
	\end{array} \right.
\end{equation}
This constitutes a PT-symmetric system similar to the thermal atomic system in the manuscript, with a linewidth of $2\kappa-4\sqrt {2{\kappa _{{\rm{in}}}}}\eta\left|a_L\right|Y_{\mathrm{in}}$ in the $\mathcal{P} \mathcal{T}$-symmetric phase. The conversion efficiency $\eta$ can adjust the strength of feedback and realize the linewidth narrowing.
\label{sec2}

\section{Experimental details}
\subsection{Derivation of the feedback factor}

\label{sec3A}

In this section, we derive the expression of feedback factor $\Gamma _{%
\mathrm{FB}}$. This factor is defined as
\begin{equation}
B_{y}=-\frac{\Gamma _{\mathrm{FB}}}{\gamma }P_{x},  \label{By1}
\end{equation}
thus we need to derive the explicit expressions of the relation between the
feedback magnetic field $B_{y}$ and the measured spin polarization $P_{x}$.

Because of the spin polarization $P_{x}$, the polarization direction of the
linearly polarized light rotates by an angle $\theta =k_{\theta }P_{x}$,
where $k_{\theta }$ is the Faraday rotation coefficient of the atoms. Then
we use the polarization homodyne detection to detect the Faraday rotation
angle $\theta $. The output voltage $V_{\mathrm{PD}}$ of the balanced
photodetector has a linear relationship with the Faraday rotation angle, $%
V_{PD}=k_{V}\theta $, where $k_{V}$ is the proportional coefficient. We
connect the output of the balanced photodetector to the feedback loop formed
by the feedback resistor $R$ and the $y$-axis feedback coil $R_{c}$ ($R\gg
R_{c}$), and the current generated is $I_{\mathrm{FB}}=V_{\mathrm{PD}}/R$,
so the feedback magnetic field is $B_{y}=k_{I}I_{\mathrm{FB}}$, where $k_{I}$
is the magnetic field generated by the coil per ampere of current. Finally,
we obtain the relationship between the feedback magnetic field $B_{y}$ and
spin polarization $P_{x}$ as
\begin{equation}
B_{y}=\frac{k_{I}k_{V}k_{\theta }}{R}P_{x}.  \label{By2}
\end{equation}%
Comparing Eq. (\ref{By1}) with Eq. (\ref{By2}), we obtain
\begin{equation}
\Gamma _{\mathrm{FB}}=-\frac{\gamma k_{I}k_{V}k_{\theta }}{R},
\end{equation}
which is inversely proportional to the feedback resistance $R$. In our
experiment, we can tune the feedback factor $\Gamma _{\mathrm{FB}}$ by
adjusting the feedback resistance $R$, with other parameters unchanged.

\subsection{Experimental parameters}
Larmor frequency $\omega_0=\gamma B_0=2\pi\times 0.35\mathrm{\ Hz/nT}\times2200\mathrm{\ nT}=2\pi\times 7.7\mathrm{\ kHz}$, relaxation time $T_1\approx0.6\mathrm{\ ms},\ T_2=0.486\mathrm{\ ms}$, delay time $\tau\approx2\ us$, optical pumping rate $R_{op}\approx1600\ \mathrm{Hz}$, equilibrium polarization $P_0\approx0.97$.
\label{sec3B}

\subsection{Energy level diagram and laser frequencies}
\label{sec3C}

\begin{figure}[h]
	\centering
	\includegraphics[width=0.6\linewidth]{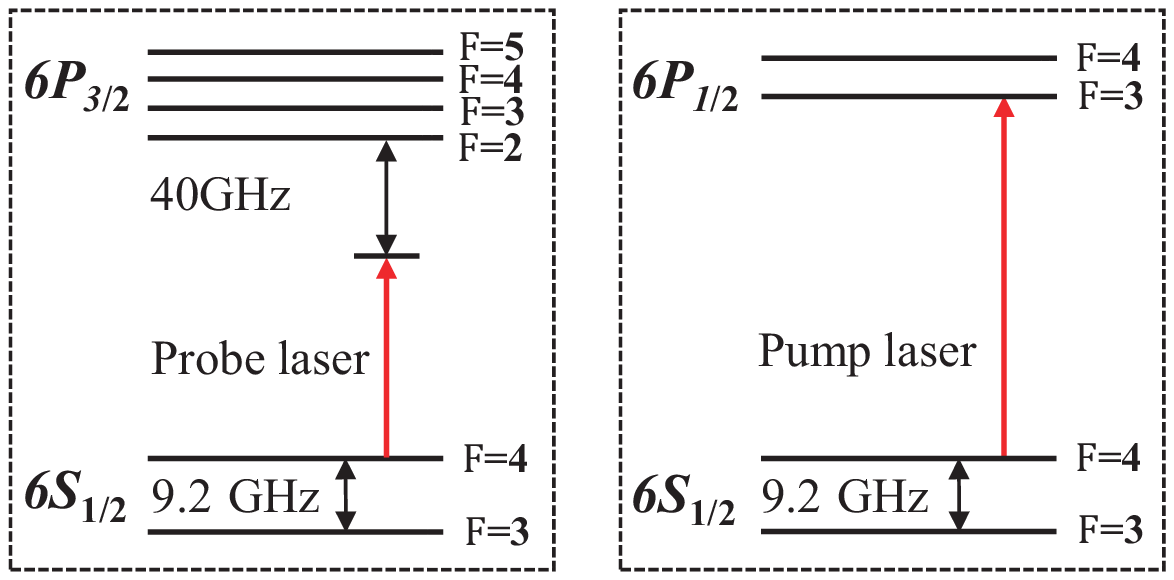}
	\caption{Energy level diagrams of $^{133}$Cs atom and the frequencies of probe 
		laser and pump laser used in the experiment.}
	\label{figS2}
\end{figure}

The energy level diagrams and laser frequencies in the experiment are shown
in Fig. \ref{figS2}. The linearly polarized probe laser power is 25 $\mu $W
and its frequency is $40$ GHz below the $F=4$ to $F^\prime =2$ transition of
the $^{133}$Cs D2 line (wavelength $852$ \textrm{nm}). The circularly
polarized pump laser power is 1 mW and its frequency is on resonance with the $F=4$ to $%
F^\prime =3$ transition of the $^{133}$Cs D1 line (wavelength $894$ \textrm{nm%
}).

\subsection{Principle of measuring the spin polarizability $P_x$}
The far detuning linearly polarized probe light incident on the atomic vapor cell along the x direction. The right and left circularly polarized components of linearly polarized light have different refractive indices due to the spin polarization $P$ of the atoms in the $x$ direction. After passing through the same distance, the phase difference between left circularly polarized light and right circularly polarized light will change, thus rotating the polarization plane of probe light. This is the Faraday rotation effect.

\begin{figure}[h]
	\centering
	\includegraphics[width=0.3\linewidth]{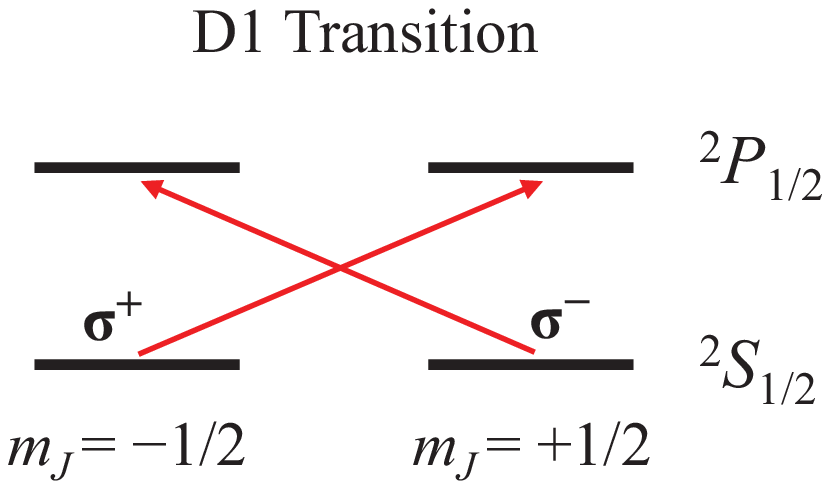}
	\caption{Energy level diagram for D1 transition of $^{133}$Cs.}
	\label{figS3}
\end{figure}
For example, we consider the D1 transition of the Cs atom in the Fig.\ref{figS3}. The ground state of cesium has an angular momentum of $+1/2$ or $-1/2$. The transition rule determines that electrons with different orientations of spin angular momentum projection will absorb photons with different polarization states. The atom with an angular momentum of $+1/2$ only absorbs $\sigma^-$ circularly polarized light, while the atom with an angular momentum of $-1/2$ only absorb $\sigma^+$ circularly polarized light. The atomic polarization (the difference in population of the ground states) makes the atomic vapor absorb light at different rates for the two different circularly polarized states, which gives rise to circular dichroism. We know that different absorbance means different refractive indices for the two circularly polarized light. This results in a phase difference between the two types of circularly polarized light emitted, which leads to a rotation of the polarization plane of the final synthesized linearly polarized light. We use a half waveplate and a polarization beam splitter to split the probe light. The intensity of the two component light fields $\left(I_1,I_2\right)$ is detected with a balanced photodetector, and then the angle of rotation of the polarization plane of the probe light is detected by subtraction $\left(I_1-I_2\right)$. Because the rotation angle of the polarization plane of the probe light is proportional to the spin polarization in the x-direction, we can obtain the information of the spin polarization $P_x$.
\label{sec3D}

\subsection{Sequence diagram of the drives for two linewidth measurement schemes}
 \begin{figure}[h]
	\centering
	\includegraphics[width=0.5\linewidth]{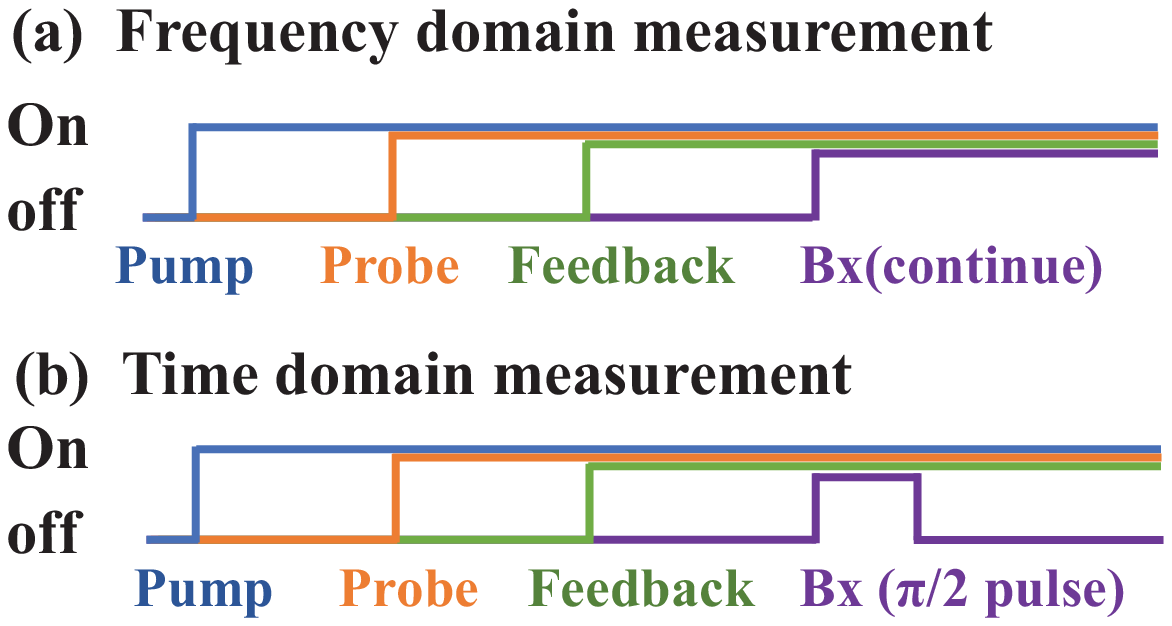}
	\caption{(a), (b) Linewidth measurement in the frequency and time domain. }
	\label{figS4}
	\end{figure}

Figure \ref{figS4} shows the sequence diagram of the drives for the two linewidth measurement schemes. The main difference is $x$-axis magnetic field $B_x$, which has different forms and different roles in the two measurement schemes.
 In the frequency domain measurement, the driving magnetic field $B_x$ is a continuous RF magnetic field, as shown in Fig.\ref{figS4} (a). We constantly change the frequency of $B_x$ and scan through the resonance frequency of spin precession. And then we use the lock-in amplifier to demodulate the resonance lineshape and obtain the resonance linewidth. In the time domain measurement, we use a pulsed magnetic field $B_x$ to rotate the atomic spins to the $x$-direction, as shown in Fig.\ref{figS4} (b). Then the spins undergo Larmor precession and gradual relaxation under the action of the $z$-direction bias magnetic field, and the linewidth can be obtained by measuring the relaxation time.

\label{sec3E}

\subsection{Discussion of measuring magnetic field and sensitivity}
The discussion of measuring magnetic field and sensitivity is divided into three parts (1) Method of measuring the magnetic field in the $z$-axis. (2) Definition of sensitivity and derivation of sensitivity formula (3) Experimental measurement procedures for signal-to-noise ratio $S/N$ and sensitivity. And the reason why the feedback method improves the sensitivity is analyzed.
\label{sec3F}
\subsubsection{Method of measuring the magnetic field in the $z$-axis}
\begin{figure}[h]
	\centering
	\includegraphics[width=0.5\linewidth]{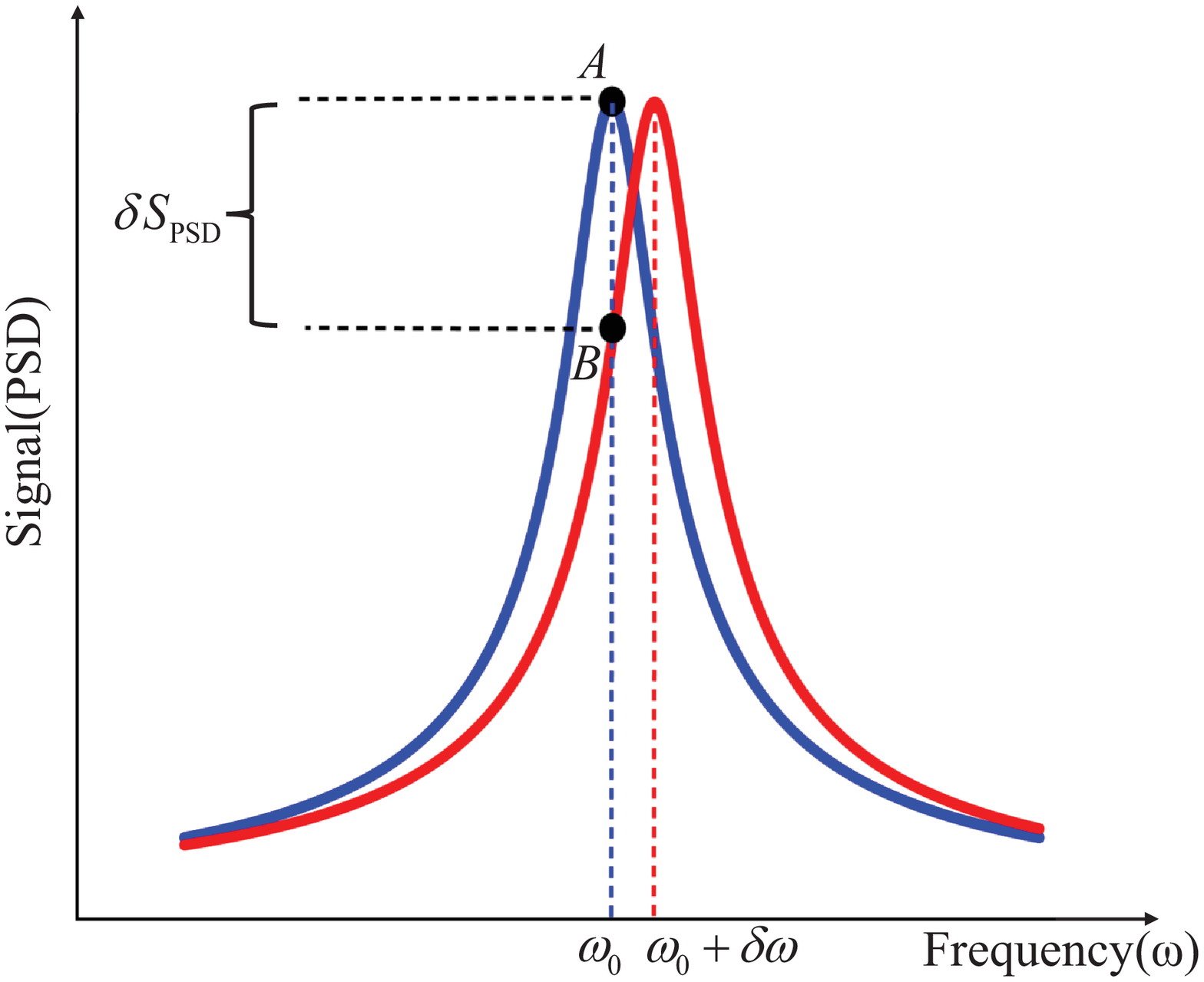}
	\caption{Schematic diagram of the power spectral density}
	\label{figS5}
\end{figure}

There is a magnetic field $B_0$ to be measured in the z-direction. We apply an $x$-axis driving field $B_x=B_1\cos{\omega t}$ and scan the frequency of the driving field $\omega$. Then, we measure the output signal of the photodetector and use the power spectral density (PSD, $S_{\mathrm{PSD}}\left(\omega\right)$) to represent the magnitude of the signal. According to the expression of spin polarization $P_x$ (Eq.\ref{Px}), the expression of signal power spectral density $S_{\mathrm{PSD}}\left(\omega\right)$ can be obtained that
\begin{equation}
	\begin{aligned}
		 S_{\mathrm{PSD}}(\omega)&=\left|P_x(\omega)\right|^2=\left(\frac{\frac{\omega_0 \omega_1 P_0}{2 \omega_{\mathrm{eff}}\left(1+\Gamma_{\mathrm{FB}} P_0 \tau\right)}\left(\omega_{\mathrm{eff}}-\omega\right)}{\left(\omega_{\mathrm{eff}}-\omega\right)^2+\frac{\Delta \omega_{\mathrm{FWHM}}^2}{4}}\right)^2+\left(\frac{\frac{\omega_0 \omega_1 P_0}{4 \omega_{\mathrm{eff}}\left(1+\Gamma_{\mathrm{FB}} P_0 \tau\right)} \Gamma_{\text {eff }}}{\left(\omega_{\mathrm{eff}}-\omega\right)^2+\frac{\Delta \omega_{\mathrm{FWHM}}^2}{4}}\right)^2 \\
		& =\frac{\left(\frac{\omega_0 \omega_1 P_0}{2 \omega_{\text {eff }}\left(1+\Gamma_{\mathrm{FB}} P_0 \tau\right)}\right)^2}{\left(\omega_{\text {eff }}-\omega\right)^2+\frac{\Delta \omega_{\mathrm{FWHM}}^2}{4}} \simeq \frac{\omega_1^2 / 4}{\left(\omega_0-\omega\right)^2+\frac{\Delta \omega_{\mathrm{FWHM}}^2}{4}}, \\
		&
	\end{aligned}
\label{eqS47}
\end{equation}
where $\omega_1=\gamma\left|B_x\right|$, $P_0\approx1$ and $\Delta\omega_{\mathrm{FWHM}}$ is the linewidth. The delay $\tau$ is small, so $\omega_{\mathrm{eff}}\approx\omega_0, 1+\Gamma_{\mathrm{FB}}P_0\tau\approx1$. As shown in Fig.\ref{figS5}, the blue curve shows the variation of the signal power spectral density $S_{\mathrm{PSD}}\left(\omega\right)$ with the frequency $\omega$ of the driving magnetic field. When the signal reaches its maximum, it means that the driving magnetic field frequency $\omega$ and the resonance center frequency $\omega_0$ are equal ($\omega=\omega_0=\gamma B_0$). According to the resonance center frequency, the magnetic field to be measured can be calculated is $B_0=\frac{\omega_0}{\gamma}$. And the max amplitude is
\begin{equation}
{S_{{\rm{PSD}}}} = \frac{{\omega _1^2}}{{\Delta \omega _{{\rm{FWHM}}}^2}}.
\label{eqS48}
\end{equation}
\label{sec3F1}
\subsubsection{Derivation of the sensitivity formula}
Suppose there is a weak change in the magnetic field to be measured $B_{\mathrm{new}}=B_0+\delta B$, then the new resonance frequency is $\omega_{\mathrm{new}}=\omega_0+\delta\omega$, and the corresponding resonance curve will be shifted, as shown in the red curve in Fig. \ref{figS5}. Because the frequency of the driving magnetic field remains constant ($\omega=\omega_0$) and is not equal to the new resonance center frequency $\omega_{\mathrm{new}}$, the signal becomes smaller, as shown by point B on the red curve. As the weak change $\delta\omega$ of the magnetic field causes the signal amplitude to change from point A to point B, the change value of the signal is $\delta S_{\mathrm{PSD}}$. When change value of the signal $\delta S_{\mathrm{PSD}}$ and noise $N_{\mathrm{PSD}}$ satisfy $\delta S_{\mathrm{PSD}}\geq N_{\mathrm{PSD}}$, we can distinguish that the signal has changed and know that the magnetic field to be measured has changed. When $\delta S_{\mathrm{PSD}}=N_{\mathrm{PSD}}$, we can just distinguish that the signal has changed, and the corresponding weak change of frequency $\delta\omega$ (magnetic field $\delta B$) is our sensitivity.
The value of the change in the signal can be written as
\begin{equation}
	\delta {S_{{\rm{PSD}}}} = \left| {\frac{{d{S_{{\rm{PSD}}}}}}{{d\omega }}} \right|\delta \omega ,
\end{equation}
When weak frequency is much small than linewidth ($\delta\omega\ll\Delta\omega_{\mathrm{FWHM}}$), we can obtain from Eq.(\ref{eqS47}) and Eq.(\ref{eqS48}) that
\begin{equation}
	\left| {\frac{{d{S_{{\rm{PSD}}}}}}{{d\omega }}} \right| = \frac{{\left( {\omega  - {\omega _0}} \right)\omega _1^2}}{{2{{\left( {\frac{{\Delta \omega _{{\rm{FWHM}}}^2}}{4} + {{\left( {\omega  - {\omega _0}} \right)}^2}} \right)}^2}}} \approx \frac{{8\omega _1^2\delta \omega }}{{\Delta \omega _{{\rm{FWHM}}}^4}} = \frac{{\omega _1^2}}{{\Delta \omega _{{\rm{FWHM}}}^2}}\frac{8}{{\Delta \omega _{{\rm{FWHM}}}^2}} = \frac{{8{S_{{\rm{PSD}}}}}}{{\Delta \omega _{{\rm{FWHM}}}^2}}.
\end{equation}
So we obtain
\begin{equation}
	\delta {S_{{\rm{PSD}}}} = \left| {\frac{{d{S_{{\rm{PSD}}}}}}{{d\omega }}} \right|\delta \omega  = \frac{{8{S_{{\rm{PSD}}}}}}{{\Delta \omega _{{\rm{FWHM}}}^2}}\delta {\omega ^2}.
\end{equation}
Using the criterion $\delta S_{\mathrm{PSD}}=N_{\mathrm{PSD}}$ for sensitivity, we obtain the expression for sensitivity $\delta\omega$ as
\begin{equation}
    \delta {S_{{\rm{PSD}}}} = \frac{{8{S_{{\rm{PSD}}}}}}{{\Delta \omega _{{\rm{FWHM}}}^2}}\delta {\omega ^2} = {N_{{\rm{PSD}}}} \to \delta \omega  = \frac{{\Delta {\omega _{{\rm{FWHM}}}}}}{{2\sqrt 2 \sqrt {{S_{{\rm{PSD}}}}/{N_{{\rm{PSD}}}}} }}.
\label{eqS52}
\end{equation}
In the actual measurement, we input the signal to the spectrum analyzer, and the measurement result is the square root of the power spectral density (i.e., $\mathrm{PS}\mathrm{D}^{1/2}$). Therefore, the relationship between the signal-to-noise ratio $S/N$ obtained in the actual measurement and the power spectral density signal-to-noise $S_{\mathrm{PSD}}/N_{\mathrm{PSD}}$ is
\begin{equation}
	\frac{S}{N} = \sqrt {\frac{{{S_{{\rm{PSD}}}}}}{{{N_{{\rm{PSD}}}}}}}.
\label{eqS53}
\end{equation}
Putting Eq.(\ref{eqS53}) into Eq.(\ref{eqS52}), we obtain the measurement sensitivity as
\begin{equation}
	\delta \omega  = \frac{1}{{2\sqrt 2 }}\frac{{\Delta {\omega _{{\rm{FWHM}}}}}}{{S/N}}.
\end{equation}
The corresponding magnetic field sensitivity is
\begin{equation}
	\delta B = {k_F}\frac{{\Delta {\omega _{{\rm{FWHM}}}}}}{\gamma }\frac{1}{{S/N}},
\label{eqS55}
\end{equation}
where $k_F=\frac{1}{2\sqrt2}$. It is worth noting that our calculation is a sensitivity analysis of the system operating very close to resonance. When the detected magnetic field change does not satisfy  $\delta\omega\ll\Delta\omega_{\mathrm{FWHM}}$, it will affect $\left|\frac{dS_{\mathrm{PSD}}}{d\omega}\right|$ and modify the coefficient $k_F$. However, the scaling rule $\delta B \propto \Delta\omega_{\mathrm{FWHM}}/(S/N) $ still holds.
\label{sec3F2}
\subsubsection{Experimental measurement procedures for signal-to-noise ratio and sensitivity}
According to Eq.(\ref{eqS55}), we need to measure the linewidth and the signal-to-noise ratio at resonance to calculate the sensitivity. The method of measuring linewidth in the time or frequency domain has been described in detail in the main text, so the following section focuses on the measurement of signal-to-noise ratio.
\begin{figure}[b]
	\centering
	\includegraphics[width=0.8\linewidth]{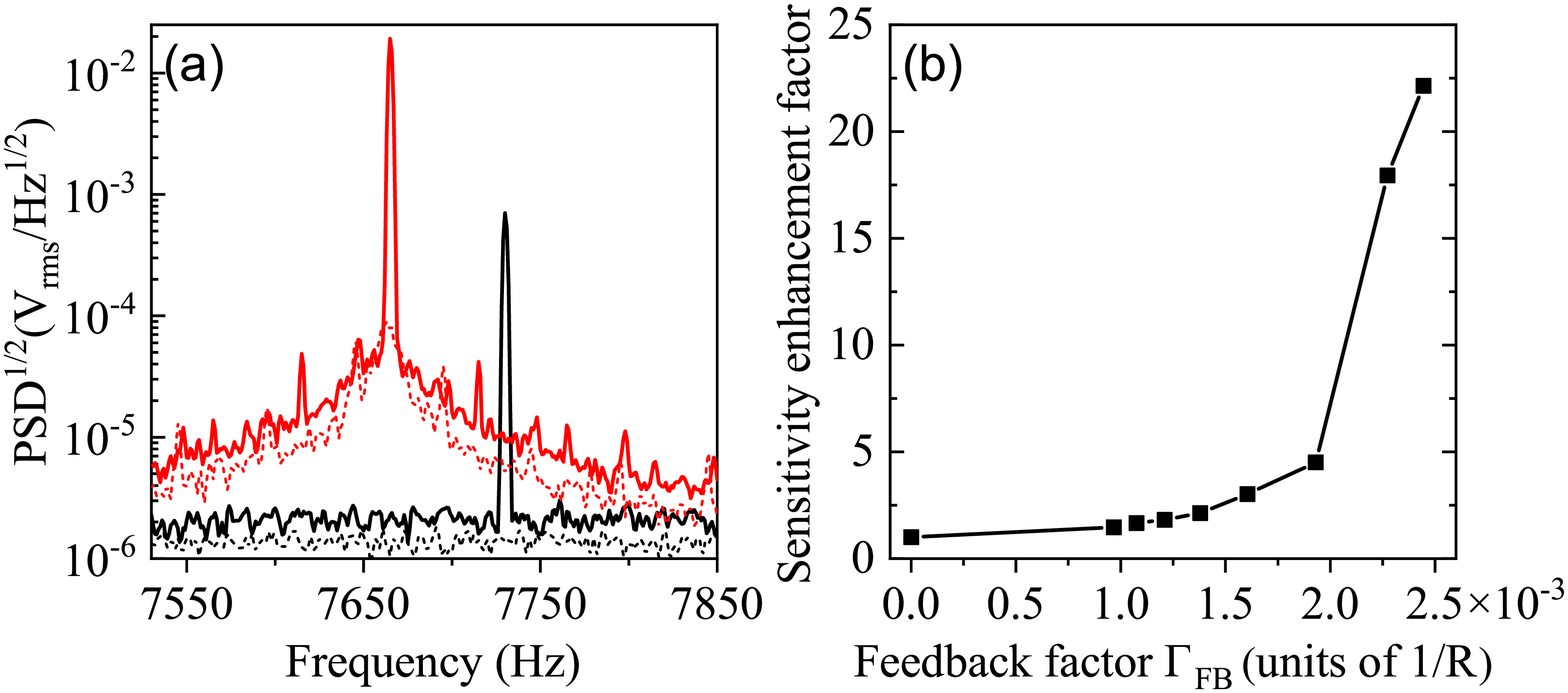}
	\caption{(a) Square root of power spectral density (PSD$^{1/2}$) for
		feedback resistance $R=409\ \Omega$ (red curves) and without feedback (black
		curves). The solid (dashed) curves are the results in the presence (absence)
		of driving magnetic field $B_{x}$. (b) Sensitivity enhancement factor of the
		$M_{x}$ magnetometer as a function of the feedback factor.}
	\label{figS6}
\end{figure}

We send the signal of the photodetector to the spectrum analyzer for measurement. First, we do not add the driving magnetic field $B_x$ to obtain the power spectral density of noise, the maximum value of the noise is $N$, as shown by the black dashed line in Fig.\ref{figS6}(a). Then, we apply the driving magnetic field $B_x$ and scan to the resonance frequency. The peak value at the resonance of the power spectral density represents the signal strength $S$, so we can obtain the signal-to-noise ratio $S/N$. With or without feedback, the signal-to-noise ratio is measured in the same way. The black curve in the Fig.\ref{figS6}(a) is the measurement result without feedback, and the red curve is the result with feedback. The signal strength with feedback (red solid line) is significantly greater than the signal strength without feedback (black solid line) but the feedback also brings an increase in noise (red dashed line), so the signal-to-noise ratio does not improve. We can measure the linewidth and signal-to-noise ratio for different feedback coefficients and calculate the sensitivity improvement factor compared to no feedback case according to Eq.(\ref{eqS55}), as shown in Fig. \ref{figS6}(b). Although the signal-to-noise ratio is not improved, the feedback scheme can narrow the linewidth by up to 48 times, which can still bring a maximum of 22 times improvement according to the formula Eq.(\ref{eqS55}).

\label{sec3F3}